# Flat-band spin density wave in twisted bilayer materials

Zhigang Song[1], Jingshan Qi[2], Olivia Liebman[3], Prineha Narang[3]

[1]John A. Paulson School of Engineering and Applied Sciences, Harvard University, Cambridge, MA 02138, USA

[2]School of Science, Tianjin University of Technology, Tianjin 300384, China

[3]College of Letters and Sciences, University of California Los Angeles, Los Angeles, CA 90095 USA

∗ To whom correspondence should be addressed: Jinshang Qi (qijingshan@email.tjut.edu.cn) or Prineha Narang (prineha@ucla.edu).

Twisting is a novel technique for creating strongly correlated effects in two-dimensional bilayered materials, and can tunably generate nontrivial topological properties, magnetism, and superconductivity. Magnetism is particularly significant as it can both compete with superconductivity and lead to the emergence of nontrivial topological states. However, the origin of magnetism in twisted structures remains a subject of controversy. Using self-developed large-scale electronic structure calculations, we propose the magnetism in these twisted bilayer systems originates from spin splitting induced by the enhanced ratio of the exchange interaction to band dispersion.

Moiré superlattices formed by twisted bilayer graphene have attracted significant interest due to their intriguing strongly correlated physics when doped at low temperature.[1, 2] At each integer electron or hole filling (except for the charge neutrality point), twisted bilayer graphene exhibits strongly correlated Chern insulator behavior. When slightly deviated away from integer filling for hole type carriers, superconductivity has been observed in twisted bilayer graphene at the magic angle.[3] Meanwhile, magnetization is a prerequisite for the presence of correlated Chern insulators and other topological and magnetic phenomena, as it breaks the time-reversal symmetry. Understanding the transition from magnetic to nonmagnetic states is crucial for revealing Cooper pairing in superconductivity in magic-angle graphene.[4] Besides, similar ferromagnetization has also been observed in other twisted bilayer materials, such as twisted bilayer $MoTe_2$ when it is quarter-filled with electrons,[5] suggesting that doping-induced magnetization is a general phenomenon in twisted materials with flat bands even beyond magic-angle graphene.

Despite numerous experimental observations,[6-8] theoretical predictions on the origin of magnetism in twisted bilayer graphene have not been well-consistent with experimental data. In twisted bilayer graphene, Chern insulators are necessarily accompanied by a nonzero integration of the Berry curvature, and this Berry curvature is predicted to induce an orbital moment.[9-12] Thus, in previous reports the ferromagnetism in twisted bilayer graphene was proposed as orbital ferromagnetism and attributed to the Berry curvature and nonlocal current.[6, 7, 11, 13-16] Orbital ferromagnetism is a new physical topic that is different from the common ferromagnetism stemmed from a local spin on an atom. Due to its novelty, orbital ferromagnetism has attracted perhaps excessive attention from experimentalists seeking to answer the question of magnetism in these twisted bilayer systems. However, several puzzles arise when experimental researchers attempt to find evidence for orbital ferromagnetism. For example, Berry curvature and nonlocal current in two-dimensional materials can induce an orbital moment that is only vertical to the material plane, but experimental measurements show that the Hall effect has a large response to an in-plane magnetic field.[7] In addition, the measured magnetic moment (~ $2\mu_B$)[17] is always less than the theoretically predicted orbital moment (5~$10\mu_B$)[10]. Although alignment with h-BN is necessary to induce a small band gap and large valley-contrast Berry curvature in graphene, ferromagnetism has also been observed without h-BN alignment.[7] Even for a large twist angle of 1.6° above the magic angle of 1.08°, ferromagnetism was also observed in experiments.[18] Besides, isospin behaviors are also thought to be closely related to magnetic properties in twisted bilayer graphene and so on.[8] These puzzles suggest that the origin of magnetism may not or at least not solely come from orbital magnetic moments, and that electron spin may be another crucial factor.

Theoretical models based on low-energy-band approximation are often limited by the difficulty of performing variational algorithms in all atomic orbitals.[19] Different possible magnetic ground states can be proposed even for the same materials, including twisted bilayer graphene.[20-23] This can lead to inconsistencies or even incorrect predictions. However, density functional theory (DFT) has the potential to shed light on these puzzles, as demonstrated by its years of successful computational predictions.[24] Due to the challenge of extremely large supercell size, the spin-polarized electronic structure study of twisted bilayer graphene has been largely unattainable until now. Through use of our self-developed code (PWmat),[25] we are uniquely capable of efficiently simulating large supercells containing 10,000 atoms or more. After optimizing the atomic orbital basis set and keeping the only necessary orbitals, the Hamiltonian matrixes can be dramatically decreased. Without significantly changing the main algorithm of DFT, we can achieve a large-scale calculation. We have confirmed the accuracy of this method in previous work by comparing it with experiments.[26-28] Our calculations use a single-zeta atomic basis set and the PBE exchange-correlation functional, with the pseudopotential of FHI applied. To include the interlayer molecular interaction, we applied DFT-D2 method,[29] and relaxed the atomic positions in the twisted structures until the force on each atom was smaller than 0.03 eV/Å. We used a 2×2×1 $k$-mesh when the twisted angle was below 1.65°, and set the temperature in Fermi smearing as 3K. Here, for the first time, we use large-scale spin-polarized DFT calculations to compute the spin response to different doping in twisted bilayer graphene and other typical twisted layers with a small twist angle. Our results show that localization-induced spin splitting and spontaneous spin density waves are a universal mechanism for the physical origin of magnetism in twisted bilayer graphene and other twisted layers beyond graphene with a small twist angle.

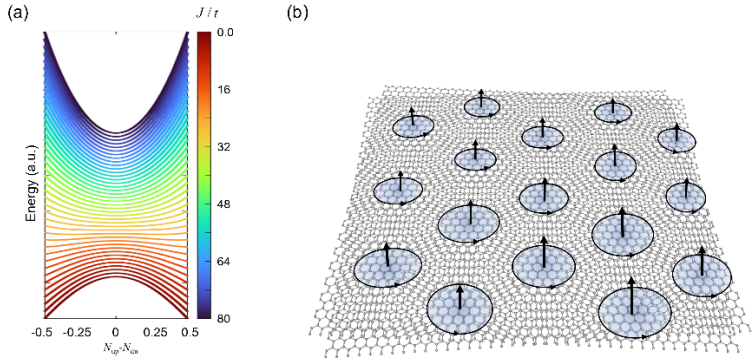

**Figure 1** (**a**) Energy as a function of occupation difference between the spin-up and spin-down bands ($N_{up}$-$N_{dn}$). Color varying from blue and red corresponds to the increase of the ratio $J$ and $t$. $t$ is set as 1. (**b**) Illustration of twist-induced collective magnetization.

We start from a toy *t-J* model, which describes the spin splitting upon doping as follows

$$H = \sum_{ij} t_{ij} c_i^\dagger c_j + h.c. + \sum_{ij} J_{ij} \mathbf{S}_i \cdot \mathbf{S}_j \tag{1}$$

where $t_{ij}$ represents the hopping integral between different orbitals on different lattice sites, and $J_{ij}$ is the effective Heisenberg exchange interaction. Under decreasing twist angle, the bands tend to flatten out as the lattice parameter increases. This band flattening implies large electron effective mass, a quenching of the electronic hopping kinetic energy, and the extended electronic states to become localized. As the twist angle decreases, both *t* and *J* decrease. Although the exact functional dependence of *t* and *J* on the twist angle is unknown, the ratio of *J/t* always increases with respect to the decreasing twisting angle.[30] We solve the above model Hamiltonian using a two-dimensional toy model of one-orbital (two-band including spin) on a square lattice. The ground states are obtained by minimizing the energy by varying the occupation of spin-up $N_{up}$ and spin-down $N_{dn}$ bands, while we keep a total occupation, which correspond to a certain doping. The energy landscape of the model is shown in Fig. 1(a). If the ratio (*J/t*) of exchange interaction to band dispersion is small, doping does not induce spin magnetization. On the other hand, when *J/t* is large, doping induces nonzero magnetization to a lower total free energy. Because these localized states are distributed in the moiré superlattice rather than on an atom, the magnetism is nonlocal and results in a spin density wave, as shown Fig. 1(b).

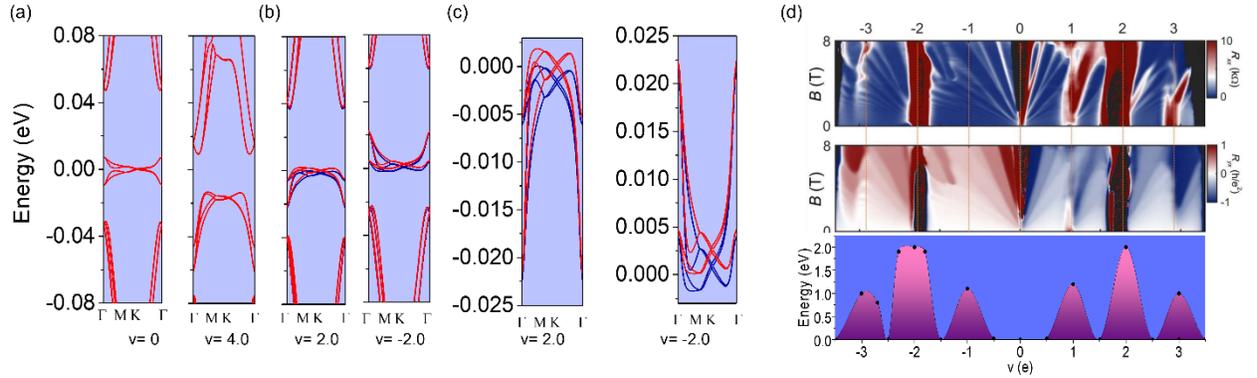

**Figure 2** (**a-b**) Spin-polarized band structures with different doping (*v*) in a supercell for twist angles of 1.47°. The Fermi level is set as 0. (**c**) Zoom in of Fig. b. Red and blue are spin-up and spin-down channels, respectively. (**d**) Comparison between the experimental phase diagram (figure from reference[31]) and calculated spin splitting upon doping.

Next, we perform spin-polarized DFT calculations of twisted bilayer graphene, with the largest system consisting of 9076 atoms. In the absence of doping, the spin-up and spin-down bands are degenerate for any twist angle. However, when the twist angle is reduced below 1.65°, the spin splitting jumps to a nonzero value upon doping with two electrons. This leads to a breaking of the degeneracy of the spin-up and spin-down bands, as in Fig. 2(a-c) and Fig. S1 for twist angles of 1.29°, 1.47° and 1.65°. As the Fermi level crosses the spin-polarized bands, the number of electrons with spin-up and spin-down is no longer equal, resulting in a net spin moment. Despite the angle of 1.65° being slightly larger than the common magic angle of 1.08° in strongly correlated physics, ferromagnetization was observed experimentally at this twist angle.[18]

To investigate the effect of doping on spin splitting and band dispersion, we focused on a twist angle of 1.47°. Fig. 1(d) shows the spin magnetic moment for different filling factors $v$. In the absence of doping, the band structure exhibits approximate particle-hole symmetry, with well-separated bands near the Fermi level. Our results show that when the number of doped electrons is less than 0.5 e/supercell, the spin splitting remains at zero. However, as the number of doped electrons increases, the spin splitting also increases and reaches a maximum average spin splitting at a doped electron number of 2 ($v$=2). After that, the spin splitting begins to decrease with further doping. At the filling factor $v= \pm 2$, the maximum spin splitting is around 2.5 meV, resulting in a maximum spin magnetic moment of 1.6 $\mu_B$ in each moiré unit cell. At the filling factors $v$=1 or 3, the spin magnetic moment is as large as 0.8 $\mu_B$ in one moiré unit cell. When the doped electron number exceeds 3.5, the spin splitting becomes zero again. When the number of doped electrons reaches 4, the Fermi level enters the band gap, turning the material into a band insulator with zero spin moment and spin splitting. The band structures with different filling factors are shown in Fig. S2. Our results suggest that the doping-induced spin splitting is likely to be the primary reason for the magnetic properties of the system, with the magnetic moment (~ 1.6$\mu_B$) caused by spin polarization being closer to the experimental measurements (~ 2$\mu_B$) in the case of $v$=2.

Another important phenomenon observed in the twisted bilayer graphene system is band renormalization, which cannot be approximated as static bands upon doping as seen in previous studies.[32] When electrons are doped into the system, the bands are bent towards the low-energy zone at the Γ point, as shown in Fig. 1(c). The bandwidth of the four bands above the Dirac point significantly decreases, while the bandwidth of the four bands below the Dirac point increases. This doping-induced bending of the bands causes a dramatic breaking of the particle-hole symmetry. Hole doping has a similar effect on the bands, but with an opposite change in the band bend and bandwidth compared to electron doping. As the doping level increases, the band renormalization monotonically increases, which is in good agreement with the measurements of local density-of-state peaks in previous experiments.[33] This finding resolves the conflict between the bandwidth from the Hamiltonian model and the experimental measurements.

A spin density wave can be shown by the spin density modulation near the Fermi level in Fig. 3. As the twist angle decreases, the electron density becomes more localized. At the twist angle of 1.47°, the electron density is almost exclusively localized in a zone of AA stacking, which is consistent with previous experimental findings.[18] The charge pattern forms a ring with a small pseudo hole at the center, where the electron density is small but not zero. The average radius of the ring is approximately 2.0 nm, indicating that the spin magnetic moment is not localized at the atomic scale. Instead, it is a collective contribution from several hundred carbon atoms and exhibits a periodic fluctuation pattern. This is in contrast to common magnetic materials, such as iron, manganese, and chromium,[25] where one spin is typically localized on a single atom.[24]

As discussed above, spin splitting is correlated with electron occupation on flat bands near the Fermi level. The electron occupation and spin splitting change coherently where if one of them is changed, the phase transition can be driven. By applying a magnetic field one can enhance spin splitting, meanwhile the band occupations can also be tunably altered by changing an external factor such as temperature. In our calculations, we include the effect of temperature in the Fermi smearing, which showed that when the temperature is above 10K, the spin splitting induced by doping disappears. Superconductivity is not yet captured in DFT calculations due to the absence of electron-electron interaction, so the phase transition to superconductivity was not captured in present calculations. With the aid of experiments, we tried to include the electron and hole occupation effects in our calculated phase diagram, shown in the lower panel of Fig. 2(d). Both the calculated spin splitting and the experimental anomalous Hall effect imply broken time-reversal symmetry, while the peak positions in doping should be the same. Indeed, the calculated phase diagram of the twisted bilayer is in good accordance with the summary from experimental measurements as shown in Fig. 2(d).

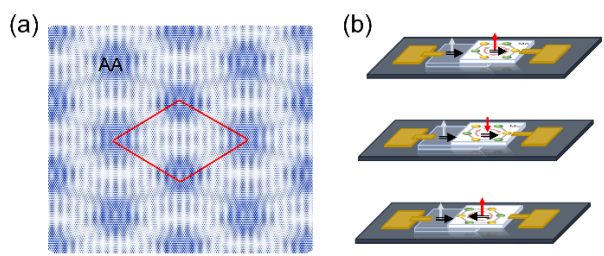

**Figure 3** (a) Spin density in the range of ±5 meV around the Fermi level under a twist angle of 1.47°. The red frame implies a moiré unit cell. (b) Illustration of proposed experiments. In a device, the left part of the heterostructure is twisted bilayer graphene, and the right part is Mn-doped $MoS_2$. Arrows represent the applied magnetic field. In-plane double arrows present the valley pseudospin.

To determine the weight of spin contribution in ferromagnetism, we propose experiments involving a spin tunneling junction consisting of Mn-doped (or Fe-doped $MoS_2$) and twisted bilayer graphene, as illustrated in Fig. 3(b). The transport in this junction is dependent on the spin configuration, and one can half-dope the twisted bilayer graphene. In previous work, monolayer $MoS_2$ with Fe or Mn doping can be ferromagnetic with a coercivity field larger than 1000 Oe,[34, 35] while the coercivity field is about 200 Oe based on the measured hysteresis loops.[6, 15] Thus, one can apply an external magnetic field in the range from 200 to 1000 Oe to switch the magnetic moment in twisted bilayer graphene only. By measuring the ratio of magneto-resistance between the spin-up and spin-down channels, we can reflect the weight of spin contribution in the ferromagnetism. If the spin does not play an important role, the magnetoresistance will be vanishingly small, whereas a large magnetoresistance indicates the significance of spin. Additionally, the transport is also dependent on the orbital magnetization. In ferromagnetic $MoS_2$ monolayers, the valley is polarized, and one can dope one valley with a net orbital magnetic moment, as shown in Fig. 3(b).[25, 36] By fixing the spin as parallel in twisted bilayer graphene and ferromagnetic $MoS_2$ and rotating the $MoS_2$ to switch the valley degree of freedom by 180°, one can determine the effect of orbital moment or Berry curvature on transport. Again, if the orbital moment is not important, the magnetoresistance will be vanishingly small, whereas a large magnetoresistance indicates the significance of orbital moment.

Guided by the toy model in Eq. (1) we also used DFT calculations to investigate the possible magnetization in other typical twisted bilayers of $h$-BN, $h$-$MoTe_2$, PbS and $h$-$NbSe_2$ in twistronics. The results are shown in Fig. 4. At larger twist angle, such as 60° in hexagonal lattices and 45° in square lattices, the moiré period is short, and wave functions are nonlocal. Thus, the spin splitting is zero. The spin splitting always increases and then decreases as the twist angle decreases. The twist angle which yields the largest spin splitting is material dependent, e.g. in twisted bilayer graphene the spin splitting peaks around 1.47°, whereas in twisted bilayer h-BN, spin splitting peaks at 6.009°. In an example of twist angle of 2.0046° for h-BN, the spin splitting is 13.5 meV, (details seen in Fig. S4). The spin polarized bands are two-fold degenerate spin-polarized states and localized in real space, as shown in Fig. S5. If the twist angle is small enough, it leads to a magnetic moment of 2 $\mu_B$, when two electrons are doped. The electronic structures in this case are similar to the states in N-V centers, so the doped h-BN could have practical application as quantum bits.[37-39] In twisted bilayer PbS, the largest spin splitting occurs roughly at a twist angle of 7.8° (details seen in Fig. S6). In a typical twist angle of 6.025°, the spin splitting is 7.1 meV and magnetic moment is 2 $\mu_B$, when two electrons are doped into the twisted bilayer PbS. [26] The spin density wave is mainly localized at AA stacking zones as shown in Fig. S7. In electron-doped twisted bilayer MoTe$_2$, the peak of spin splitting is at 9.43°. Band structures under small twist angles and spin density waves are shown in Fig. S8 and Fig. S9, respectively. In experiments, the ferromagnetism has been observed in twisted bilayer MoTe$_2$ with a twist angle around 4°, when it is doped.[5]

According to the *t-J* model, doping is not necessary if the ratio of *J/t* is large enough in a metallic twisted bilayer. Here we consider the effects of twist on spin polarization of undoped hexagonal NbSe$_2$, which without twisting is a nonmagnetic metal. In the larger twisted angle above 9.43°, there is still no spin polarization; yet as the twist angle decreases, the spin splitting increases. When the twist angle is larger than 4.4085°, spin splitting can be as large as 0.6 eV (see Fig. 4(d) and Fig. S10). The spin polarization in a moiré super-cell is above 20 $\mu_B$. As shown in Fig. S11, the spin density wave is partially localized, and a large proportion is extended beyond a moiré cell.

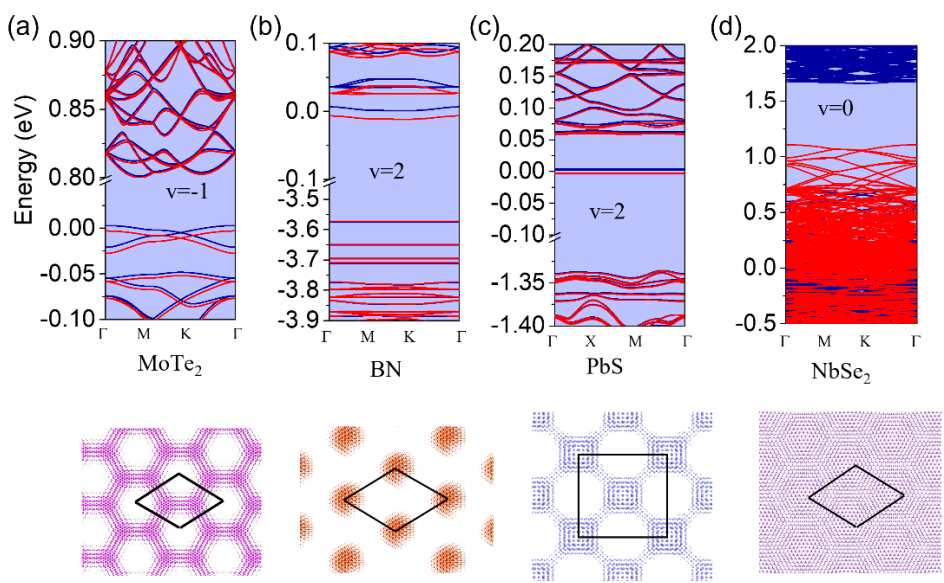

**Figure 4** **Band structures of some typical twisted bilayer materials.** (**a**) Bilayer MoTe$_2$ twisted by 4.4085°. (**b**) h-BN bilayer twisted by 2.0046°. (**c**) PbS bilayer twisted by 6.0256°. (**d**) NbSe$_2$ bilayer twisted by 4.4085°. The Fermi level is set as 0. Lower panels are corresponding spin density near the Fermi level.

In conclusion, our DFT calculations propose a different mechanism for the origin of magnetism in twisted bilayer graphene, which is consistent with previous experiments, especially the cascade of phase transitions upon doping. Our results suggest spin density wave order accompanied by flat bands may be the driving force of these phase transitions, but we do not exclude the possibility of an important contribution from the orbital magnetization. Our calculations also show that the doping-induced magnetism in flat bands is not unique to twisted bilayer graphene, but can be observed in other twisted systems as well. Especially in some nonmagnetic metallic bilayers, such as bilayer $NbSe_2$, twisting can transform a nonmagnetic state into a ferromagnetic state by tuning the localization and correlation of electron motion. The good agreement between our ultra-large-scale spin-polarized DFT calculations and experimental observations paves the way for high-throughput calculations in twisted bilayer materials.

**Acknowledgements**

This work is supported by the supported by the Quantum Science Center (QSC), a National Quantum Information Science Research Center of the U.S. Department of Energy (DOE). P.N. gratefully acknowledges support from the Gordon and Betty Moore Foundation grant number #8048 and from the John Simon Guggenheim Memorial Foundation (Guggenheim Fellowship).

# Supporting Information

# Flat-band spin density wave in twisted bilayer materials

Zhigang Song[1], Jingshan Qi[2], Olivia Liebman[3], Prineha Narang[3]

[1]John A. Paulson School of Engineering and Applied Sciences, Harvard University, Cambridge, MA 02138, USA

[2]School of Science, Tianjin University of Technology, Tianjin 300384, China

[3]College of Letters and Sciences, University of California Los Angeles, Los Angeles, CA 90095 USA

∗ To whom correspondence should be addressed: Jinshang Qi (qijingshan@email.tjut.edu.cn) or Prineha Narang (prineha@ucla.edu).

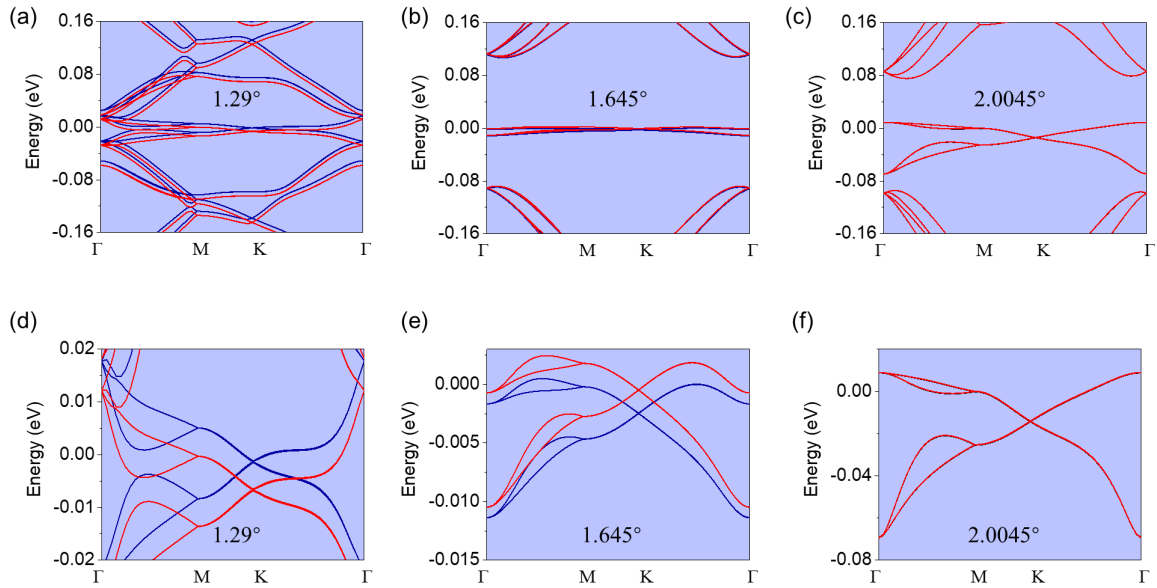

**Figure S1** **(a-c)** Spin-polarized band structures under different twist angles. The filling factor is $v$=2. **(d-f)** Zoom in on Fig. **a-c** near the Fermi level. Red and blue are spin-up and spin-down channels, respectively.

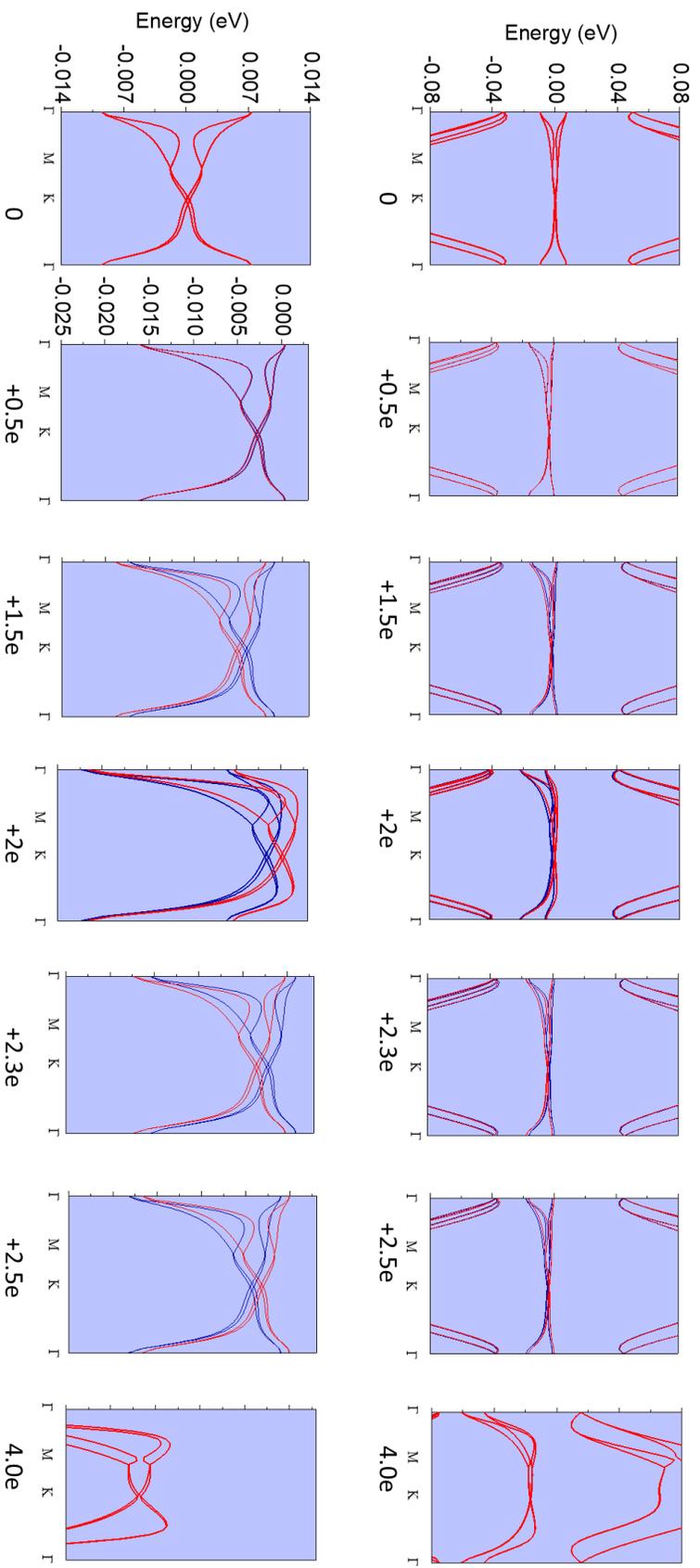

**Figure S2** Spin-polarized band structures of twisted bilayer graphene with different filling factor. Lower row is the corresponding zoom in the upper row near the Fermi level.

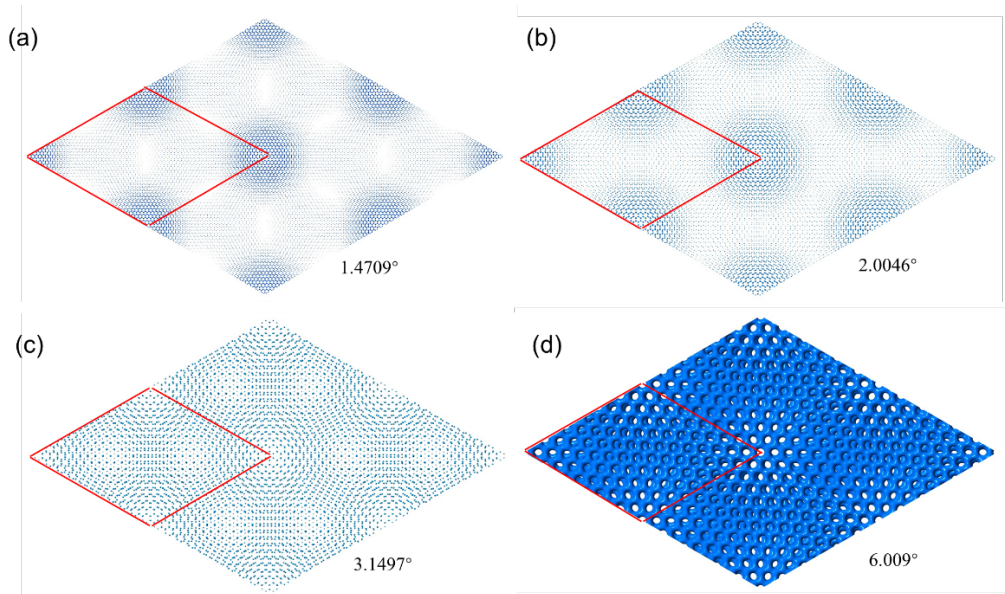

**Figure S3** **(a-d)** Charge density of twisted bilayer graphene in the range from ±5 meV near the Fermi level under different twist angles, given above. Red frame implies the moiré unit cell. In Fig. a, the charge density is approximately equal to spin density. In Figs. b-d, there is no spin polarization.

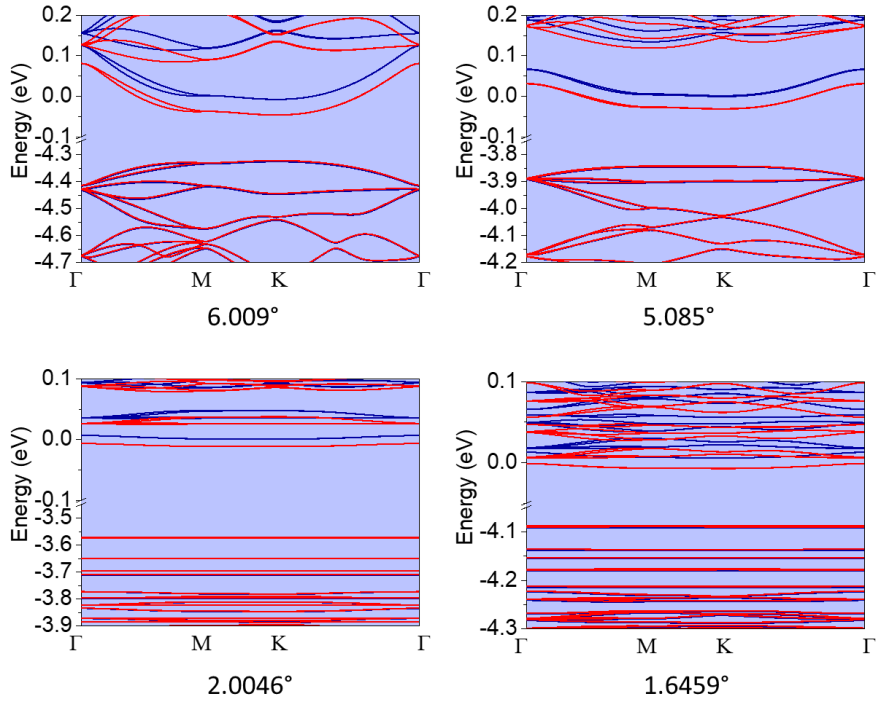

**Figure S4 Spin-polarized band structures of twisted bilayer h-BN under different twist angles.** The filling factor is *v*=2.

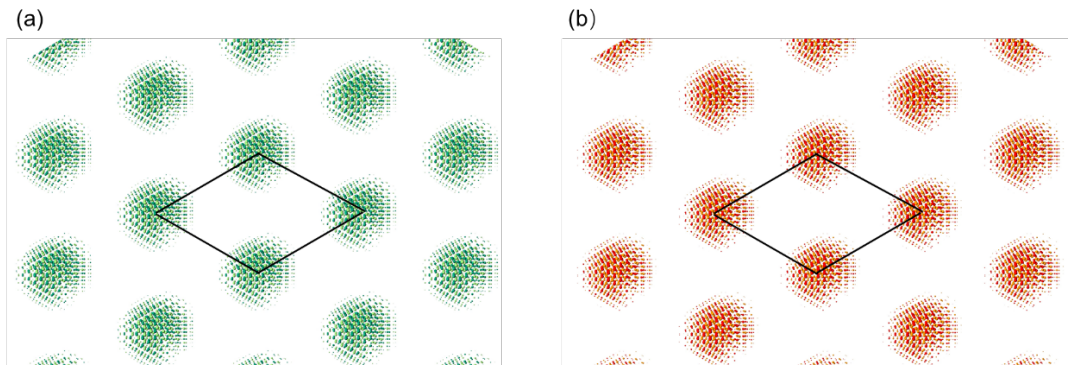

**Figure S5 Spin-polarized wave functions of twisted bilayer h-BN.** The filling factor is *v*=2. Red and orange are Bloch wave functions of two degenerate bands below the Fermi level. Twist angle is 2.0045°.

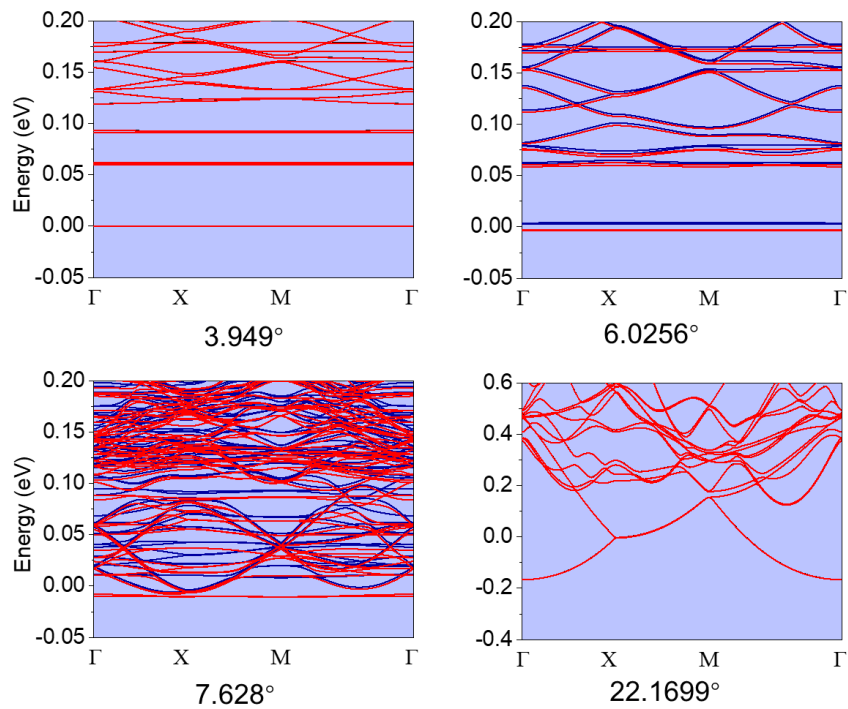

**Figure S6 Spin-polarized band structures of twisted bilayer PbS under different twist angles.** The filling factor is $v=2$.

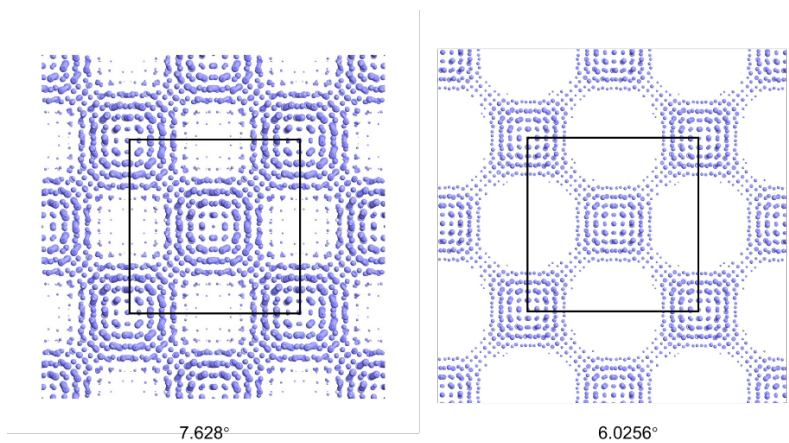

**Figure S7 Spin density of twisted bilayer PbS.** The filling factor is $v=2$. The charge density is cut off in the range from -0.2 eV to 0.

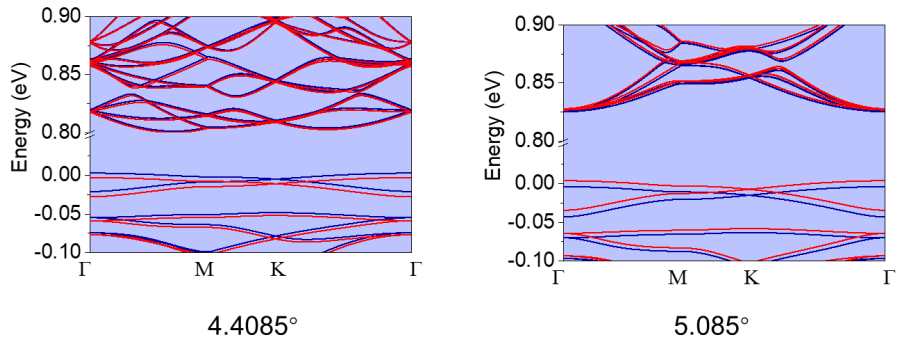

**Figure S8 Spin-polarized band structures of twisted bilayer MoTe$_2$ under different twist angles.** The filling factor is $v=-1$.

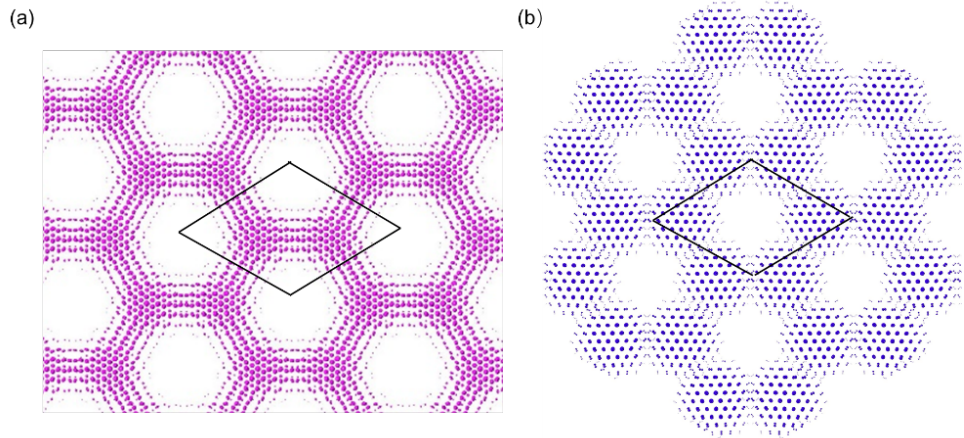

**Figure S9 Spin density of twisted bilayer MoTe$_2$.** **(a)** The filling factor is $v=-1$. The charge density is cut off in the range from 0 to 0.1. **(b)** The filling factor is $v=1$. The charge density is cut off in the range from -0.2 eV to 0. Twist angle is 4.4085°.

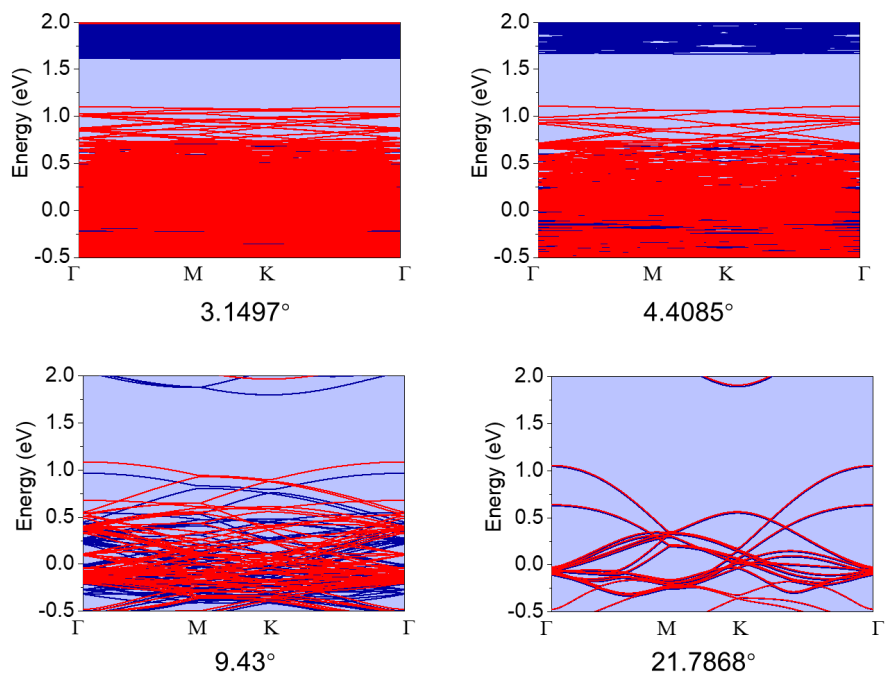

**Figure S10 Spin-polarized band structures of twisted bilayer NbSe$_2$ under different twist angles.** The filling factor is *v*=0. No doping is applied.

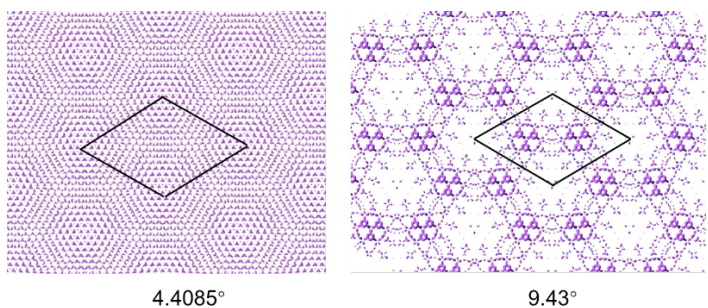

**Figure S11 Spin density of twisted bilayer NbSe$_2$.** The spin density is cut off in the range from -0.3 to 0.3eV. No doping is applied. The spin density under twist angle of 21.78° is zero.